\documentclass[aps,prl,preprint,groupedaddress,showkeys]{revtex4}

\usepackage{graphicx}
\usepackage{amsmath}
\usepackage{amsfonts}
\usepackage{amssymb}
\usepackage{graphicx}
\usepackage{enumerate}

\begin{document}

\title{Bayesian networks for enterprise risk assessment}

\author{C. E. Bonafede $^{\dagger}$, P. Giudici $^{\dagger\dagger}$}
\email[$^{\dagger}$]{bonafede@eco.unipv.it}
\email[$^{\dagger\dagger}$]{giudici@unipv.it}
\homepage{www.datamininglab.it}
\affiliation{University of Pavia}

\keywords{Bayesian Networks, Enterprise Risk Assessment, Mutual Information}

\date{\today}

\begin{abstract}
According to different typologies of activity and priority, risks can assume diverse meanings and it can be assessed in different ways.\\
\indent In general risk is measured in terms of a probability combination of an event (frequency) and its consequence (impact). To estimate the frequency and the impact (severity) historical data or expert opinions (either qualitative or quantitative data) are used. Moreover qualitative data must be converted in numerical values to be used in the model.\\
\indent In the case of enterprise risk assessment the considered risks are, for instance, strategic, operational, legal and of image, which many times are difficult to be quantified. So in most cases only expert data, gathered by scorecard approaches, are available for risk analysis.\\
\indent The Bayesian Network is a useful tool to integrate different information and in particular to study the risk's joint distribution by using data collected from experts.\\
\indent In this paper we want to show a possible approach for building a Bayesian networks in the particular case in which only prior probabilities of node states and marginal correlations between nodes are available, and when the variables have only two states.
\end{abstract}

\maketitle


\section{Introduction}

A Bayesian Net (BN) is a directed acyclic graph (probabilistic expert system) in which every node represents a random variable with a discrete or continuous state \cite{2,3}.\\
\indent The relationships among variables, pointed out by arcs, are interpreted in terms of conditional probabilities according to Bayes theorem.\\
\indent With the BN is implemented the concept of conditional independence that allows the factorization of the joint probability, through the Markov property, in a series of local terms that describe the relationships among variables:

\begin{center}
$f(x_1,x_2,...,x_n) = \prod_{i=1}^n f(x_i|pa(x_i))$
\end{center}

\noindent where $pa(x_i)$ denotes the states of the predecessors (parents) of the variable $X_i$ (child) \cite{1, 2, 3, 6}. This factorization enable us to study the network locally.\\     
\indent A Bayesian Network requires an appropriate database to extract the conditional probabilities (parameter learning problem) and the network structure (structural learning problem) \cite{1,3,13,16}.\\
\indent The objective is to find the net that best approximates the joint probabilities and the dependencies among variables.\\ 
\indent After we have constructed the network one of the common goal of bayesian network is the probabilistic inference to estimate the state probabilities of nodes given the knowledge of the values of others nodes. The inference can be done from children to parents (this is called diagnosis) or vice versa from parents to children (this is called prediction) \cite{2,13,15}.\\     
\indent However in many cases the data are not available because the examined events can be new, rare, complex or little understood. In such conditions experts' opinions are used for collecting information that will be translated in conditional probability values or in a certain joint or prior distribution (Probability Elicitation) \cite{11,12,16,19}.\\
\indent Such problems are more evident in the case in which the expert is requested to define too many conditional probabilities due to the number of the variable's parents. So, when possible, is worthwhile to reduce the number of probabilities to be specified by assuming some relationships that impose bonds on the interactions between parents and children as for example the noisy-OR and its variation and genralization \cite{3,9,10,14,16}.\\
\indent In the business field, Bayesian Nets are a useful tool for a multivariate and integrated analysis of the risks, for their monitoring and for the evaluation of intervention strategies (by decision graph) for their mitigation  \cite{3,5,7}.\\
\indent Enterprise risk can be defined as the possibility that something with an impact on the objectives happens, and it is measured in terms of combination of probability of an event (frequency) and of its consequence (impact).\\
\indent The enterprise risk assessment is a part of  Enterprise Risk Management (ERM) where to estimate the frequency and the impact distributions historical data as well as expert opinions are typically used \cite{4,5,7,6}. Then such distributions are combined to get the loss distribution.\\
\indent In this context Bayesian Nets are a useful tool to integrate historical data with those coming from experts which can be qualitative or quantitative \cite{19}.


\section{Our proposal}
What we present in this work is the construction of a Bayesian Net for having an integrated view of the risks involved in the building of an important structure in Italy, where the risk frequencies and impacts were collected by an ERM procedure unsing expert opinions. \\
\indent We have constructed the network by using an already existing database (DB) where the available information are the risks with their frequencies, impacts and correlation among them. In total there are about 300 risks.\\      
\indent In our work we have considered only the frequencies of risks and no impacts. With our BN we construct the risks' joint probability and the impacts could be used in a later phase of scenario analysis to evaluate the loss distribution under the different scenarios \cite{5}.\\
\indent In table 1 there is the DB structure used for network learing and in which each risk is considered as a binary variable (one if the risk exists {\itshape (yes)} and zero if the risk dosen't exist {\itshape (not)}). Therefore, for each considered risk in the network there will be one node with two states ($one \equiv Y$ and $zero \equiv N$).

\begin{table}[htp]
\caption{Expert values database structure (Learning table)}
\label{tab:1}
\centering
{\scriptsize
		\begin{tabular}{|c|c|c|c|c|}
		\hline
    \bfseries PARENT&\bfseries CHILD&\bfseries CORRELATION&\bfseries PARENT FREQ.&\bfseries CHILD FREQ.\\
    \hline
    RISK A & RISK B & $\rho_{AB}=0.5$ & {\itshape P(risk A = Yes)}=0.85 & {\itshape P(risk B = Yes)}=0.35 \\
    \hline
    RISK A & RISK C & $\rho_{AC}=0.3$ & {\itshape P(risk A = Yes)}=0.85 & {\itshape P(risk C = Yes)}=0.55 \\  	
    \hline
		\end{tabular}}
\end{table} 

\indent The task is, therefore, to find the conditional probabilities tables by using only the correlations and the marginal frequencies. Instead, the net structure is obtained from table 1 by following the node relationships given by correlations.\\
\indent The main ideas for finding a way to construct a BN have been: first to find the joint probabilities as functions of only the correlations and the marginal probabilities; second to understand how the correlations are linked with the incremental ratios or the derivatives of the child's probabilities as functions of the parent's probabilities. This choice is due to the fact that parent and child interact through the values of conditional probabilities; the derivatives are directly linked to such probabilities and, therefore, to the degree of interaction between the two nodes and, hence with the correlation.\\
\indent Afterwards we have understood as to create equations, for the case with dependent parents we have used the local network topology to set the equations.\\
\indent We have been able to calculate the CPT up to three parents for each child. Although there is the possibility to generalize to more than three parents, it is necessary to have more data which are not available in our DB. So when four or more parents are present we have decided to divide and reduce to cases with no more than three parents. To approximate the network we have ``separated'' the nodes that give the same effects on the child (as for example the same correlations) by using auxiliary nodes \cite{3}. When there was more than one possible scheme available we have used the mutual information (MI) criterion as a discriminating index by selecting the approximation with the highest total MI; this is the same to choose the structure with the minimum distance between the network and the target distribution \cite{17,18}.\\
\indent We have analyzed first the case with only one parent to understand the framework, then it has been seen what happens with two independent parents and then dependent. Finally we have used the analogies between the cases with one and two parents for setting the equations for three parents.


\subsection{One parent case solution}

The case with one parent (figure 1) is the simplest. Let P(F) and P(C) be the marginal probability given from expert (as in table 1): 

\begin{itemize}
	\item For the parent, F, we have: P(F=Y)=x, P(F=N)=1-x; 
	\item For the child, C, we have: P(C=Y)=y, P(C=N)=1-y;
\end{itemize}

\begin{figure}[htbp]
	\centering
		\includegraphics[scale=0.65]{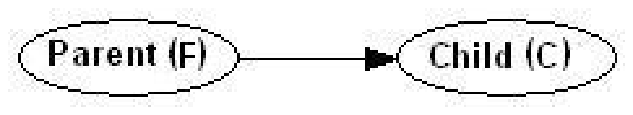}
	\caption{\label{fig:1}One parent scheme}
\end{figure}

\indent The equations to find either the conditional probabilities or the joint probabilities are:

\begin{table}[hp]
\centering
\scriptsize
		\begin{tabular}{l||l}
		  {\itshape CPT equation system}&{\itshape Joint equation system}\\
		  $\alpha_1x + \alpha_2(1-x)=y;$&$c_1=\rho M+xy;$\\
			$\alpha_1  - \alpha_2=k;$&$c_2=y-\rho M-xy;$\\
			$\alpha_1 + \alpha_3=1;$&$c_3=x-\rho M-xy;$\\
			$\alpha_2 + \alpha_4=1;$&$c_4=1-y-x+\rho M+xy;$\\   
		\end{tabular}	
\end{table}

\noindent where $k=\rho \sqrt{\frac{Var[C]}{Var[F]}}$; whit $\alpha_{i}$ and $c_{i}$ we indicate respectively the conditional and the joint probabilities.\\
\indent Considering that probabilities $c_i$ and $\alpha_{i}$ must be positive either the marginal probabilities or the correlation value should be constrained. If the marginal probabilities are fixed then the correlation values must be constrained, which will be normally the case, as estimates of probabilities are more reliable.\\
\indent It is not possible to have any value of correlation given the marginal probabilities. Indeed, as we want to maintain the marginal probabilities as fixed by the expert, correlation limits can be determined as follows:

\begin{center}
	$\rho >-\frac{xy}{M}=A;\: \rho > \frac{y+x(1-y)-1}{M}=D;\: \rho < \frac{x(1-y)}{M}=B;\: \rho < \frac{y(1-x)}{M}=C;$
\end{center}

\noindent and the correlation interval will be:

$$\rho \in \left[max(A,D);min(B,C)\right];$$


\subsection{Two parents case solutions}
This case (figure 2) is more complicated than the one before. In this situation a further difficulty is that the given expert correlations are only pairwise marginal and, therefore, we need more information to find the CPT.\\
\indent For example the joint moment among the nodes which is not in the DB. Consequently there can be more than one CPT, corresponding to different values of the joint moment, for the same marginal correlation and probability. 

\begin{figure}[hbtp]
	 		\centering
			\includegraphics[scale=0.6]{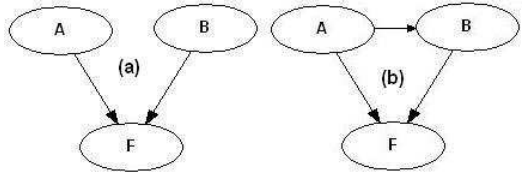}
			\caption{Two independent parents (a) and dependent parents (b)}
			\label{fig:2}
\end{figure}

\indent The joint moment becomes thus a project parameter to be set by using an appropriate criterion. We define the standardized joint moment among three variables to be:

\begin{center}
$\rho_{N_iN_jN_k}=\frac{E[(N_i-E[N_i])(N_j-E[N_j])(N_k-E[N_k])]}{\sqrt[3]{Var[N_i]Var[N_j]Var[N_k]}}$.
\end{center}

\indent To choose among such CPTs we have used the total mutual information ($I_{total}$) by selecting that CPT with the $\rho_{N_iN_jN_k}$ that gives the minimum $I_{total}$.\\
\indent In this case we have to distinguish between independent and dependent parents. The solutions are:

\begin{table}[hp]
\centering
\scriptsize
		\begin{tabular}{l||l}
{\itshape CPT equation system for independent parents}&{\itshape Joint equation system for dependent parents}\\
$f(x,z)=(\alpha_1-\alpha_2-\alpha_3+\alpha_4)xz+(\alpha_2-\alpha_4)x+$ & $c_1=\rho_{ABF}M_{ABF}+xyz+(\rho_{AF}M_{AF})z+$\\
$+(\alpha_3-\alpha_4)z+\alpha_4=y$& $+(\rho_{BF}M_{BF})x+(\rho_{AB}M_{AB})y$\\
$\frac{\partial f}{\partial x}=(\alpha_1-\alpha_2-\alpha_3+\alpha_4)z+(\alpha_2-\alpha_4)=\frac{(\rho_{AF})(M_{AF})}{x(1-x)}$&$c_1+c_5=\rho_{BF}M_{BF}+zy$\\ 
$\frac{\partial f}{\partial z}=(\alpha_1-\alpha_2-\alpha_3+\alpha_4)x+(\alpha_3-\alpha_4)=\frac{(\rho_{BF})(M_{BF})}{z(1-z)}$& $c_1+c_3=\rho_{AF}M_{AF}+xy$\\
$\frac{\partial^2 f}{\partial x\partial z}=\frac{\partial^2 f}{\partial z\partial x}=(\alpha_1-\alpha_2-\alpha_3+\alpha_4)=\frac{(\rho_{ABF})(M_{ABF})}{x(1-x)z(1-z)}$& $c_1+c_2=\rho_{AB}M_{AB}+xz$\\ 
$\alpha_1+\alpha_5=1$& $c_1+c_2+c_3+c_4=x$\\
$\alpha_2+\alpha_6=1$& $c_1+c_5+c_6+c_2=z$\\
$\alpha_3+\alpha_7=1$& $c_1+c_3+c_5+c_7=y$\\
$\alpha_4+\alpha_8=1$& $c_8=1-\sum_{i=1}^7c_i$\\    
		\end{tabular}	
\end{table}

\noindent where $M_{BF}=\sqrt{z(1-z)y(1-y)}$, $M_{AF}=\sqrt{x(1-x)y(1-y)}$, $M_{AB}=\sqrt{x(1-x)z(1-z)}$ and $M_{ABF}=\sqrt[3]{x(1-x)z(1-z)y(1-y)}$. As before the $\alpha_{i}$ and $c_{i}$ are respectively the conditional and the joint probabilities.\\
\indent The problem is now setting the marginal correlations when those given from experts are not consistent with the marginal probabilities. Differently from the case with one parent where the correlation belongs to an interval, with two parents the admissible pairs $(\rho_{AF}, \rho_{BF})$ can be shown to belong to an area.\\
\indent To approach this problem we have decided to decrease the values of the two correlations $\rho_{BF}$ and $\rho_{AF}$ with a fixed step by maintaining their relative difference. At each step we verified the existence of a value of $\rho_{ABF}$ which supports the new pair $(\rho_{AF}, \rho_{BF})$. If it exists the process is stopped, otherwise it goes to the next step; and so on.\\
\indent If the correlation $\rho_{AB}$ is different from zero (dependent parents), we can set it in advance using the interval obtained for the case of one parent; afterward the $\rho_{AB}$'s value is used into the joint equation system. Then we can work again only on the pair $(\rho_{AF}, \rho_{BF})$ by considering the same procedure for independent parents and selecting $\rho_{N_iN_jN_k}$.


\subsection{Three parents case solutions}
As before two equation systems are obtained. One system for the case with independent (see figure 3 a) parents by which the CPT is directly calculated; another one when there are some correlations between parents (see figure 3 b) and in this case the joint probabilities are calculated instead of the conditional ones. To define the equation systems the analogies between the cases with one and two parents have been exploited.

\begin{figure}[htp]
	\centering
		\includegraphics[scale=0.5]{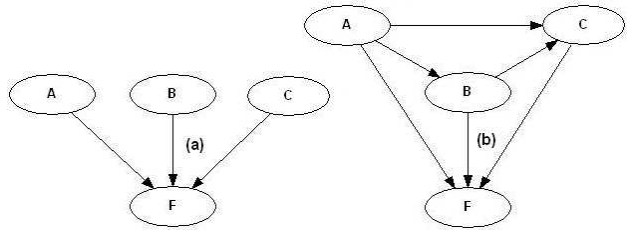}
	\caption{Three independent parents (a) and dependent parents (b).}
	\label{fig:3}
\end{figure}

\indent The solutions for independent and dependent parents are in table 2. In such equations, obviously, there will be more missing data which are all the standardized joint moments among every two parents and the child and among all parents and the child. So what we do in such a situation is to use the procedure for the case of two parents for each pair of nodes and set the correlation values such that they will be feasible for the all couples. Note that the correlation levels are now less than in previous cases. Moreover in this case the standardized joint moment among all variables is set at zero to make the research less complex.\\
\indent Furthermore, difficulties arise when there are large differences among the parents' marginal probabilities. Therefore, when there are more than three parents, we have decided to split them. Parents are split from the others by using the mutual information criterion \cite{17,18}.\\
\indent As before, for the case of dependent parents to select the feasible marginal correlations and the standardized joint moments, we start to look for the admissible correlation between the nodes with one parent (A and B), then for the nodes with two parents (C has B and A as predecessor) and finally we set the joint moment and marginal correlations for the node with three parents (F). Obviously, now the procedure is more complex and it is more difficult to select the parameters.

\begin{table}[hp]
\caption{\label{tab:2} Equation systems for three parents scheme}
\centering
\scriptsize

		\begin{tabular}{l||l}
{\itshape CPT equation system for independent parents}&{\itshape Joint equation system for dependent parents}\\
$f(x,z,w)=(\alpha_1-\alpha_2-\alpha_3+\alpha_4-\alpha_5+\alpha_6+\alpha_7-\alpha_8)xzw+$& $c_1=\rho_{ABCF}M_{ABCF}+xyzw+\rho_{ABC}M_{ABC}y+\rho_{ABF}M_{ABF}w+$\\
$+(\alpha_2-\alpha_4-\alpha_6+\alpha_8)xz+(\alpha_3-\alpha_4-\alpha_7+\alpha_8)wz+$ & $+\rho_{ACF}M_{ACF}z+\rho_{BCF}M_{BCF}x+\rho_{AB}M_{AB}wy+$\\
$+(\alpha_5-\alpha_6-\alpha_7+\alpha_8)wz+(\alpha_6-\alpha_8)z+(\alpha_4-\alpha_8)x+$ & $+\rho_{AC}M_{AC}zy+\rho_{BC}M_{BC}xy+\rho_{AF}M_{AF}xy+$\\
$+(\alpha_7-\alpha_8)w+\alpha_8=y$ & $ +\rho_{BF}M_{BF}xw+\rho_{CF}M_{CF}xz$\\
$\frac{\partial f}{\partial x}=\frac{(\rho_{AF})\sqrt{x(1-x)y(1-y)}}{x(1-x)}$ & $c_1+c_2+c_3+c_4+c_5+c_6+c_7+c_8=w$ \\
$\frac{\partial f}{\partial z}=\frac{(\rho_{BF})\sqrt{z(1-z)y(1-y)}}{z(1-z)}$ & $c_1+c_2+c_3+c_4+c_9+c_{10}+c_{11}+c_{12}=x$\\ 
$\frac{\partial f}{\partial w}=\frac{(\rho_{CF})\sqrt{w(1-w)y(1-y)}}{w(1-w)}$ & $c_1+c_2+c_5+c_6+c_9+c_{10}+c_{13}+c_{14}=z$\\
$\frac{\partial^2 f}{\partial x\partial z}=\frac{(\rho_{ABF})\sqrt[3]{x(1-x)z(1-z)y(1-y)}}{x(1-x)z(1-z)}$ & $c_1+c_3+c_5+c_7+c_9+c_{11}+c_{13}+c_{15}=y$\\
$\frac{\partial^2 f}{\partial w\partial z}=\frac{(\rho_{ABF})\sqrt[3]{w(1-w)z(1-z)y(1-y)}}{w(1-w)z(1-z)}$ & $c_1+c_2=\rho_{ABC}M_{ABC}+xyz+\rho_{AB}M_{AB}w+\rho_{AC}M_{AC}z+\rho_{BC}M_{BC}x$\\
$\frac{\partial^2 f}{\partial x\partial w}=\frac{(\rho_{ABF})\sqrt[3]{w(1-w)x(1-x)y(1-y)}}{x(1-x)w(1-w)}$ & $c_1+c_9=\rho_{ABF}M_{ABF}+xyz+\rho_{AF}M_{AF}z+\rho_{BF}M_{BF}x+\rho_{AB}M_{AB}y$\\
$\frac{\partial^3 f}{\partial x\partial w\partial z}=\frac{(\rho_{ABCF})\sqrt[4]{w(1-w)z(1-z)y(1-y)x(1-x)}}{x(1-x)z(1-z)w(1-w)}$ & $c_1+c_3=\rho_{ACF}M_{ACF}+xyz+\rho_{AF}M_{AF}w+\rho_{CF}M_{CF}x+\rho_{AC}M_{AC}y$\\
$\alpha_1+\alpha_9=1$ & $c_1+c_5=\rho_{BCF}M_{BCF}+xyz+\rho_{BF}M_{BF}w+\rho_{CF}M_{CF}z+\rho_{BC}M_{BC}y$\\
$\alpha_2+\alpha_{10}=1$ & $c_1+c_2+c_9+c_{10}=\rho_{AB}M_{AB}+xz$\\
$\alpha_3+\alpha_{11}=1$ & $c_1+c_2+c_3+c_4=\rho_{AC}M_{AC}+xw$\\
$\alpha_4+\alpha_{12}=1$ & $c_1+c_3+c_9+c_{11}=\rho_{AF}M_{AF}+xy$\\
$\alpha_5+\alpha_{13}=1$ & $c_1+c_2+c_5+c_6=\rho_{BC}M_{BC}+zw$\\
$\alpha_6+\alpha_{14}=1$ & $c_1+c_3+c_5+c_7=\rho_{CF}M_{CF}+wy$\\
$\alpha_7+\alpha_{15}=1$ & $c_1+c_5+c_9+c_{13}=\rho_{BF}M_{BF}+zy$\\
$\alpha_8+\alpha_{16}=1$ & $c_{16}=1-\sum^{15}_{i}c_i$\\
		\end{tabular}	
\end{table}


\section{Conclusion}
So far we have seen that using the equation systems for conditional and joint probabilities the CPTs can be obtained. The method can be generalized to the case with more three parents, but there are problems in setting more parameters (standardized joint moment) and in looking for more complicated feasible marginal correlation areas.\\
\indent So to develop a network we propose to use, separately, firstly the equations and procedure for the one parent; secondly those for two parents distinguishing when they are dependent and not. Finally we use the equations and the procedures, when possible, for the three parents case by distinguishing also in this situation between dependent and independent parents; otherwise we split one parent from the others by using the mutual information as splitting index \cite{17,18}.\\
\indent We remark that we need to reduce to a more simple case those configurations with more than three parents. We can achieve this trying to estimate a local approximate structure, with only one, two and three parents, by "separating" those that give different effects on the child (as for instance different incremental ratios). If there are more schemes available for the substitution we select that with the highest MI ($I_{total}$) \cite{17,18}.\\
\indent It is important to be noted that such method is modular, this is if we add or delete a node we can use the appropiate system (one, two or three nodes) to according to we add or delete a parent or a child.

\section{Acknowledgments}
The authors acknowledge financial support from the MIUR-FIRB 2006-2009 project and MUSING project contract number 027097.

\end{document}